# Strong Magnon-Magnon Coupling in Synthetic Antiferromagnets

Changting Dai and Fusheng Ma[*]

Jiangsu Key Laboratory of Opto-Electronic Technology, Center for Quantum Transport and Thermal Energy Science, School of Physics and Technology, Nanjing Normal University, Nanjing 210023, China

*Corresponding author. E-mail: phymafs@njnu.edu.cn

## ABSTRACT

Synthetic antiferromagnet, comprised of two ferromagnetic layers separated by a non-magnetic layer, possesses two uniform precession resonance modes: in-phase acoustic mode and out-of-phase optic mode. In this work, we theoretically and numerically demonstrated the strong coupling between acoustic and optic magnon modes. The strong coupling is attributed to the symmetry breaking of the system, which can be realized by tilting the bias field or constructing an asymmetrical synthetic antiferromagnet. It is found that the coupling strength can be highly adjusted by tuning the tilting angle of bias field, the magnitude of antiferromagnetic interlayer exchange coupling, and the thicknesses of ferromagnetic layers. Furthermore, the coupling between acoustic and optic magnon modes can even reach the ultrastrong coupling regime. Our findings show high promise for investigating quantum phenomenon with a magnonic platform.

Quantum magnonics,[1–3] the coherent interaction between magnons and the elementary excitations of matter, has attracted widespread attention owing to the advantage of low power dissipation in quantum information processing and storage. Recently, the strong magnon-magnon coupling[4–15] has been reported in a single magnetic nanostructure[7,8,10,16-18] and magnetic metal-insulator hybrid nanostructure.[4–6,9,11,13] Theses findings are particularly appealing, since the magnonic cavity has smaller volume and lower Ohmic losses than microwave cavity.[6,9] While, these two types of structure possess weak intralayer or interlayer exchange interaction, which lead to the coupling strength between magnon modes is relatively small, and the reported maximum value is usually around 1 GHz. From an application point of view, stronger coupling is expected since it ensures the magnons can exchange energy several times with preserved coherence before they are consumed. It is found that the magnon-magnon coupling can be highly enhanced with the presence of stronger intralayer exchange interaction in a compensated ferrimagnet gadolinium iron garnet.[8] The maximum coupling strength can approach 6.38 GHz at the coupling field of 650 mT and the temperature of 282 K.

In order to investigate the materials with interlayer exchange interaction, the synthetic antiferromagnet (SAF) has become the preferred choice due to the adjustable strength of interlayer exchange coupling (IEC) between two ferromagnetic (FM) layers.[19–21] And it is a useful system for antiferromagnetic spintronics,[22–26] its potential applications are conducive to engineer high-speed devices and improve the capabilities of spintronic devices. The magnetization dynamics of SAF are characterized by two uniform precession magnon modes, *i.e.* in-phase acoustic mode (AM) and out-of-phase optic mode (OM),[27–33] whose frequencies can reach the X and K bands. The coupling between AM and OM of SAF has been demonstrated in CoFeB/Ru/CoFeB.[12,14] The obtained maximum coupling strength is about 1 GHz, and such a weak coupling is due to the small effective IEC. Thus, it is worth to investigate the impact of the effective IEC of SAFs on the strength of magnon-magnon coupling.

In this work, we theoretically and numerically studied the strong magnon-magnon coupling in both symmetrical and asymmetrical SAFs. By breaking the symmetry of SAFs, the repulsive coupling between in-phase AM and out-of-phase OM is realized. For symmetrical SAFs, the bias field is tilted toward out-of-plane to break the symmetry of system, and the coupling strength can be tuned by adjusting the tilting angle of bias field, the magnitude of IEC, and the thicknesses of FM layers. The obtained maximum coupling strength is ~6.90 GHz, which is greater than the dissipation rates of coupled magnon modes approaching the strong coupling regime. Actually, the corresponding coupling rate is 0.36 and reaches the ultrastrong coupling regime. Furthermore, for asymmetrical SAFs, the structural difference between the thicknesses of the two FM layers can also break the intrinsic symmetry and results in the strong magnon-magnon coupling even without tilting the bias field.

Typical SAF multilayer is depicted in Fig. 1(a). The static magnetizations of the bottom and top FM layers, $M_1$ and $M_2$, have three relative distributions depending on the strength of the bias field: antiparallel, spin-canted,[34] and parallel. The $M_1$ and $M_2$ is: at the antiparallel state when $H_{ext} < H_{sc}$; at the spin-canted state at intermediate fields $H_{sc} < H_{ext} < H_s$; and at the parallel state when $H_{ext} > H_s$. The precession-phase relations of the in-phase AM and the out-of-phase OM are shown in Fig. 1(b).

For the symmetrical SAFs with same thicknesses $t_1 = t_2$, the resonance frequencies $f_{AM}$ and $f_{OM}$ of AM and OM can be described as follows when the bias field $H_{ext}$ is applied in the film plane[12,35]

$$f_{AM} = \frac{\mu_0 \gamma}{2\pi} \sqrt{(H_{ext}\cos\theta_0)(H_{ext}\cos\theta_0 - H_{IEC}\cos2\theta_0 + M_s + H_{IEC})}, \quad (1)$$

$$f_{OM} = \frac{\mu_0 \gamma}{2\pi} \sqrt{(H_{ext}\cos\theta_0 - 2H_{IEC}\cos2\theta_0)(H_{ext}\cos\theta_0 - H_{IEC}\cos2\theta_0 + M_s - H_{IEC})}, \quad (2)$$

where $H_{IEC}$ is the effective field of interlayer exchange coupling, and it can also be obtained from the following equation

$$H_{IEC} = H_s/2, \quad (3)$$

where $H_s$ is the critical field that saturate $M_1$ and $M_2$ for $\phi = 0°$. Under different magnetic fields, the angles between the equilibrium directions of $M_1$, $M_2$ and $H_{ext}$ are $\theta_1$ and $\theta_2$. Since the symmetrical SAFs,

$$\theta_1 = \theta_2 = \theta_0 = \begin{cases} \cos^{-1}\dfrac{H_{\text{ext}}}{2H_{\text{IEC}}}, & (H_{\text{sc}} < H_{\text{ext}} < H_{\text{s}}) \\ 0, & (H_{\text{ext}} \geq H_{\text{s}}) \end{cases}. \quad (4)$$

Moreover, the orientations of $M_1$ and $M_2$ are parallel to the bias field in the saturation phase, and the AM is in-phase procession, thus this mode would behave as the ferromagnetic resonance mode of a single thin film and can be described by Kittel formula[36]

$$f_{\text{FMR}} = \frac{\mu_0 \gamma}{2\pi} \sqrt{H_{\text{ext}}(H_{\text{ext}} + M_{\text{s}})}. \quad (5)$$

The dependences of the frequencies of the two modes on the bias field is shown in Fig. 1(c), in which an intersection is formed between the AM and OM in the spin-canted state. Once the symmetry of SAFs is broken by tilting $H_{\text{ext}}$ from $x$-$y$ plane toward $z$-axis, the coupling between the AM and OM will occur, forming an anti-crossing gap, as expected by the inset in Fig. 1(c). To describe the coupled resonance modes in symmetrical SAFs, the matrix formalism of the Landau-Lifshitz-Gilbert (LLG) equation was solved, whose solutions capture the features of coupled magnon modes.[7] Here, the solutions of LLG equation are equivalent to solving the following two-by-two matrix:

$$\begin{vmatrix} \omega_{\text{AM}}^2(H_{\text{ext}}, \phi) - \omega^2 & \Delta^2(H_{\text{ext}}, \phi) \\ \Delta^2(H_{\text{ext}}, \phi) & \omega_{\text{OM}}^2(H_{\text{ext}}, \phi) - \omega^2 \end{vmatrix} = 0, \quad (6)$$

where the angular frequency of bare AM

$$\omega_{\text{AM}} = \mu_0 \gamma (H_{\text{ext}} \cos\phi) \sqrt{1 + (M_{\text{s}}/2H_{\text{IEC}})}, \quad (7)$$

and the angular frequency of bare OM

$$\omega_{\text{OM}} = \mu_0 \gamma (2H_{\text{IEC}} M_{\text{s}}[1 - (H_{\text{ext}}^2/H_{\text{FM}}^2)] + \{(\sin^2\phi)/[1 + (M_{\text{s}}/2H_{\text{IEC}})]^2\} H_{\text{ext}}^2)^{1/2}. \quad (8)$$

The $H_{\text{FM}}$ is the field that can saturate the magnetizations $M_1$ and $M_2$ completely along the orientation of $H_{\text{ext}}$ at an out-of-plane angle $\phi$, its value is

$$1/H_{\text{FM}}^2 = \cos^2\phi/(2H_{\text{IEC}})^2 + \sin^2\phi/(2H_{\text{IEC}} + M_{\text{s}})^2, \quad (9)$$

and $H_{\text{FM}} = H_{\text{s}} = 2H_{\text{IEC}}$ at $\phi = 0°$. The term $\Delta$ represents the coupling of AM and OM,

$$\Delta = \mu_0 \gamma H_{\text{ext}} [2H_{\text{IEC}}/(2H_{\text{IEC}} + M_{\text{s}}) \sin^2\phi \cos^2\phi]^{1/4}. \quad (10)$$

Therefore, the coupling between these two modes in the symmetrical SAFs can be tuned by adjusting the effective IEC field $H_{IEC}$ and the out-of-plane angle $\phi$ of the bias field. The $H_{IEC}$, $H_{IEC} = |J_{IEC}|/(M_s t)$, is dependent on the thickness $t_1$ and $t_2$, saturation magnetization $M_s$, and the strength of IEC $J_{IEC}$.

The micromagnetic simulations were performed by using Mumax[3].[37] The type and strength of the IEC between two FM layers are determined by the sign and magnitude of $J_{IEC}$. Positive (negative) value represents ferromagnetic (antiferromagnetic) IEC. Here, we set a negative $J_{IEC}$. Each FM layer is discretized into cubic cells with a size of $4 \times 4 \times C_z$ nm$^3$ (along $x$, $y$ and $z$-axis). During the simulations, both the length and the width of SAFs are fixed at 400 nm, while the thicknesses are varied as indicated. For the symmetrical SAFs, $C_z = t_1 = t_2$, and for the asymmetric SAF, $C_z = 0.15$ nm. The material parameters used in the simulation are [Co/Ni/Co]/Ru/[Co/Ni/Co],[38] those are saturation magnetization $M_s = 6 \times 10^5$ A/m, exchange stiffness constant $A_{ex} = 1.3 \times 10^{-11}$ J/m, gyromagnetic ratio $\gamma/2\pi = 2.8 \times 10^{-2}$ GHz/mT, and damping constant $\alpha = 0.01$. The initial magnetization of the bottom and top FM layers are set to be antiparallel and parallel to the $x$-axis, respectively. Firstly, the magnetizations of the SAFs were stabilized by an external bias field applied at an angle, $\phi$, from the film plane. Secondly, to excite the magnons with frequencies ranging from 0 to 100 GHz, a spatially uniform sinc-function-type perturbation field $h_{rf}$, $h_{rf}(t) = h_0 \sin(2\pi f t)/(2\pi f t)$,[39,40] with the amplitude $h_0 = 5$ mT and the cut-off frequency $f = 100$ GHz was applied along the $x$ or $y$-axis. In order to excite AM and OM, two perturbation configurations were used: the transverse configuration ($h_{rf}$ is normal to $H_{ext}$) and the longitudinal configuration ($h_{rf}$ is parallel to $H_{ext}$). The time evolution of spatially averaged magnetization $m(t)$ is recorded and processed by Fourier transform to obtain the magnon spectra at the specific bias field.

Firstly, we studied a symmetrical SAF with $t_1 = t_2 = 0.3$ nm, $J_{IEC} = -0.3 \times 10^{-4}$ J/m$^2$. The theoretically calculated results are shown in Fig. 1(c) for the bias field $H_{ext}$ applied along the $x$-axis. In the spin-canted state for $H_{sc} < H_{ext} < H_s$, the dispersions of the mode ① and the mode ②, corresponding to AM and OM, display a crossing where is expected to exhibit the presence of anti-crossing depending on the coupling between the two modes. In the parallel state for $H_{ext} > H_s$, the high frequency mode ③ is the Kittel mode, while the low frequency mode ④ is the OM. However, the mode ④ is not experimentally detectable as the net magnetization is zero as shown in Fig. 1(b). The numerically simulated results are shown in Figs. 1(d) and (e) for the transverse and longitudinal perturbation configurations, respectively. For the transverse perturbation configuration as shown in Fig. 1(d), $h_{rf}$ along the y-axis can only excite the AM when $M_1$ and $M_2$ is in the spin-canted state and the Kittel mode when $M_1$ and $M_2$ is in the parallel state. This can be explained by a net magnetization between $M_1$ and $M_2$ along the direction of $h_{rf}$ as depicted in Fig. 1(b). In the spin-canted state, the net magnetization of the in-phase AM is non-zero, and which of the Kittel mode in the parallel state is also non-zero. However, the net magnetization of the out-of-phase OM is zero and hence it cannot be excited. For the longitudinal perturbation configuration as shown in Fig. 1(e), $h_{rf}$ along the x-axis can only excite the OM in the spin-canted state. Since the net magnetization of both the in-phase AM and Kittel mode along the direction of $h_{rf}$ is zero.

To excite the AM and the OM simultaneously, we tilted the $H_{ext}$ from x-axis toward y-axis by an angle $\varphi = 45°$ in the x-y plane, and keep the perturbation field $h_{rf}$ along the y-axis. Figure 2(a) is the color-plot of the frequency-field ($f$-$H_{ext}$) dispersion and clearly displays a crossing between the AM and the OM in the spin-canted state. This can be understood as that the tilted $H_{ext}$ has two components, which are perpendicular and parallel to the $h_{rf}$, respectively. Thus, this tilted perturbation configuration can combine the results of both the transverse and longitudinal configurations as shown in Figs. 1(d) and (e). Moreover, for $\varphi = 45°$, the tilted perturbation configuration is equally sensitive to the AM and OM, which is reflected in the similar intensity of both modes in the dispersion. The theoretical calculated results according to Eqs. (1) and (2) are also shown as green dashed lines in Fig. 2(a), which is in good agreement with the simulated results. It suggests that the in-plane deflected bias field only affect the sensitivities of both modes, but it is unable to break the symmetry of the symmetrical SAFs without the presence of an anti-crossing at the crossing point of the dispersions of the AM and the OM.

From Figs. 1(c) and 2(a), we found that the frequency of AM is lower than that of OM when $H_{ext} < H_I$, where $H_I$ denotes the magnetic field of the intersection of the AM and OM, while the AM possesses the higher frequency when $H_I < H_{ext} < H_s$. It can be understood as that, when $H_{ext} < H_I$, the angle between $M_1$ and $M_2$ is $\theta_1 + \theta_2 > 90°$, the $M_1$ and $M_2$ tend to be antiparallel for the AM, which is supported by the IEC energy, so the AM has lower frequency. In contrast, the $M_1$ and $M_2$ tend to deviate from the antiparallel configuration for the OM, so the OM needs to work against the IEC energy and possesses higher frequency. When $H_I < H_{ext} < H_s$, the angle between $M_1$ and $M_2$ is $\theta_1 + \theta_2 < 90°$, the in-phase AM is unfavorable by the IEC energy compared with the out-of-phase OM, thus the AM has higher frequency than that of the OM. The intersection of the AM and OM, i.e. $H_{ext} = H_I$, implies the angle of $M_1$ and $M_2$ is $\theta_1 + \theta_2 = 90°$. However, it should be noted that the symmetrical SAF are protected by the intrinsic symmetry for the in-plane applied bias fields, as a result there is no anti-crossing observed in the $f$-$H_{ext}$ dispersions. On the other hand, by breaking the symmetry of SAFs, we expect to acquire the anti-crossing in the $f$-$H_{ext}$ dispersions of the AM and OM as shown in the inset of Fig. 1(c).

One of the effective methods to break the symmetry is to employ an out-of-plane component of the bias field.[7,14] Here, we tilt the $H_{ext}$ from $x$-axis toward $z$-axis by an angle $\phi$ in the $x$-$z$ plane, and keep the perturbation field $h_{rf}$ along the $y$-axis. Taking the symmetrical SAF with $t_1 = t_2 = 0.3$ nm, $J_{IEC} = -0.3 \times 10^{-4}$ J/m$^2$ as an example, the dispersions of various magnons at $\phi = 30°$, 60°, and 90° ($\varphi = 0°$ unless otherwise noted) are shown in Figs. 2(b)-(d), respectively. A pronounced anti-crossing between the previously uncoupled resonant modes occurred for $\phi = 30°$ and 60°. The presence of anti-crossing is attributed to the hybridization of AM and OM, i.e. leading to the magnon-magnon coupling. While for $\phi = 90°$, the SAF has the symmetry around the out-of-plane orientation, the magnon-magnon anti-crossing will vanish again as shown in Fig. 2(d). At the coupling field $H_g$, where the coupled modes have similar intensities, we named them as "up mode" (high frequency branch) and "down mode" (low frequency branch), respectively. To certify the interaction between the AM and the OM, we define the coupling strength $g$ as the half of the modes splitting frequency at $H_g$. For the up and down modes at $H_g$, their dissipation rates $k_{up}$ and $k_{down}$ are defined as the half width at half maximum of the line broadenings. For $\phi = 30°$, the coupling strength $g = |f_{up} - f_{down}|/2 = 2.80$ GHz, where $f_{up}$ and $f_{down}$ refer to the frequencies of the up and down modes at $H_g$, and the dissipation rates are $k_{up} = 0.58$ GHz and $k_{down} = 0.54$ GHz at $H_g = 650$ mT. Thus, the condition $g > k_{up}, k_{down}$ is satisfied and the magnon-magnon coupling reaches the strong coupling regime.[41] Moreover, the theoretically calculated results based on Eq. (6) are depicted by green dashed lines in Figs. 2(b)-(d) in the spin-canted state. The theoretical calculations can roughly reproduce the characteristics of the coupled magnon modes. Figure 2(e) shows the frequencies of the up and down modes as a function of the tilting angle $\phi$ for the SAF. The splitting between the up and down modes increases with $\phi$. It is worth to point out that the strong coupling between magnons are realized in the X or K band, which is not possible for the magnons in the ferromagnets.

We now turn to investigate the factors affecting the coupling strength. As shown in Fig. 2(f), $g$ increases with increasing $\phi$ when the thicknesses of FM layers are unchanged, $t_1 = t_2 = 0.3$ nm, and $J_{IEC}$ is fixed. Furthermore, $g$ increases with increasing the magnitude of $J_{IEC}$, such as $g$ increases from 1.80 to 6.90 GHz when $|J_{IEC}|$ increases from $0.1 \times 10^{-4}$ to $0.3 \times 10^{-4}$ J/m$^2$ with fixing $t_1 = t_2 = 0.3$ nm and $\phi = 60°$ unchanged. In addition, we also introduce a unitless parameter, coupling rate $g/f_g$, to describe the extent of coupling, where $f_g = (f_{up} + f_{down})/2$ is the coupling center frequency. The color column charts display the dependences of $g/f_g$ on $\phi$ for three different $J_{IEC}$. For $g/f_g > 0.1$, it is usually refer to the condition of realizing the ultrastrong coupling regime,[8,42,43] *i.e.* the coupling strength can be compared with the resonance frequency. This requirement is obviously satisfied for $\phi > 30°$ and $|J_{IEC}| > 0.2 \times 10^{-4}$ J/m$^2$ according to the horizontal guide line $g/f_g = 0.1$ in Fig. 2(f).

On the other hand, we intend to investigate the influence of thicknesses $t_1 = t_2 = t$ of FM layers of symmetrical SAFs on the coupling strength $g$. Figures 3(a)-(c) show the simulated and calculated $f$-$H_{ext}$ dispersion of magnons for the symmetrical SAFs with $J_{IEC}$ = -0.3 × 10$^{-4}$ J/m$^2$, $\phi$ = 60° and $t$ = 0.3, 0.4, and 0.45 nm, respectively. The theoretical calculations are according to Eq. (6), and which is in consistence with the simulated features. For $t$ = 0.3/0.4/0.45 nm, $g$ = 6.90/3.65/2.65 GHz, $k_{up}$ = 0.64/0.51/0.59 GHz, and $k_{down}$ = 0.55/0.38/0.33 GHz at $H_g$ = 800/600/530 mT, thus these magnon-magnon couplings reach the strong regime. As shown in Fig. 3(d), $g$ increases with decreasing $t$ when $J_{IEC}$ = -0.3 × 10$^{-4}$ J/m$^2$ is unchanged, and $\phi$ is fixed. Furthermore, $g$ increases with increasing the $\phi$, such as $g$ increases from 0 GHz to 6.90 GHz when $\phi$ increases from 0° to 60° with fixing $J_{IEC}$ = -0.3 × 10$^{-4}$ J/m$^2$ and $t$ = 0.3 nm unchanged. The color column charts display the dependences of $g/f_g$ on $t$ for three different $\phi$. The requirement of ultrastrong coupling is obviously satisfied for $t$ < 0.35 nm and $\phi$ > 30° according to the horizontal guide line $g/f_g$ = 0.1 in Fig. 3(d).

So far, we have investigated the magnon-magnon coupling in symmetrical SAF with $t_1$ = $t_2$ = $t$. Actually, if $t_1 \neq t_2$, the intrinsic symmetry of the SAFs can also be broken. To verify whether the structural asymmetry can cause the coupling between the AM and OM in the spin-canted state, we simulated an asymmetrical SAF with $t_1$ = 0.45 nm for bottom FM layer, $t_2$ = 0.3 nm for top FM layer, and $J_{IEC}$ = -0.3 × 10$^{-4}$ J/m$^2$. In order to compare with the symmetrical SAFs, we apply an in-plane bias field $\phi$ = 0° and keep perturbation field along $y$-axis. Figure 4(a) is the dispersion of a symmetrical SAF with $t_1$ = $t_2$ = 0.3 nm, $J_{IEC}$ = -0.3 × 10$^{-4}$ J/m$^2$, $\phi$ = 0° and $\varphi$ = 45°. In contrast, as shown in Fig. 4(b), the simulated $f$-$H_{ext}$ dispersion of the asymmetrical SAF exhibits an obvious anti-crossing in the spin-canted state. The coupling strength $g$ = 5.49 GHz, and the dissipation rates are $k_{up}$ = 0.86 GHz, $k_{down}$ = 0.71 GHz at the coupling field $H_g$ = 950 mT, which satisfies $g$ > $k_{up}$, $k_{down}$, thus reaches the strong coupling regime. It also belongs to the ultrastrong magnon-magnon coupling since the $g/f_g$ = 0.19 is greater than 0.1. Therefore, the structurally asymmetrical SAF can provide a platform for realizing the strong AM-OM coupling under an in-plane bias field. Furthermore, employing different $M_s$ for bottom and top FM layers is also possible to induce the breaking of the intrinsic symmetry of the SAFs.[7]

In summary, we theoretically and numerically demonstrated the strong coupling between acoustic and optic magnon modes in both the symmetrical and asymmetrical synthetic antiferromagnets. In order to realize the magnon-magnon coupling, two effective methods are employed to break the symmetry of system: applying an out-of-plane bias field and constructing a structurally asymmetrical SAF. We found that the coupling strength is proportional to the tilting angle of the bias field from the film plane and the magnitude of interlayer exchange coupling, while is inversely proportional to the thicknesses of ferromagnetic layers. Furthermore, the system can even reach the ultrastrong coupling regime. Our findings provide a possibility to produce and tune the strong magnon-magnon coupling in synthetic antiferromagnets.

This work was supported by the National Natural Science Foundation of China (Grant Nos. 12074189 and 11704191), the Natural Science Foundation of Jiangsu Province of China (Grant No. BK20171026), the Jiangsu Specially-Appointed Professor, and the Six-Talent Peaks Project in Jiangsu Province of China (Grant No. XYDXX-038).

**DATA AVAILABILITY**

The data that support the findings of this study are available from the corresponding author upon reasonable request.

**Figures**

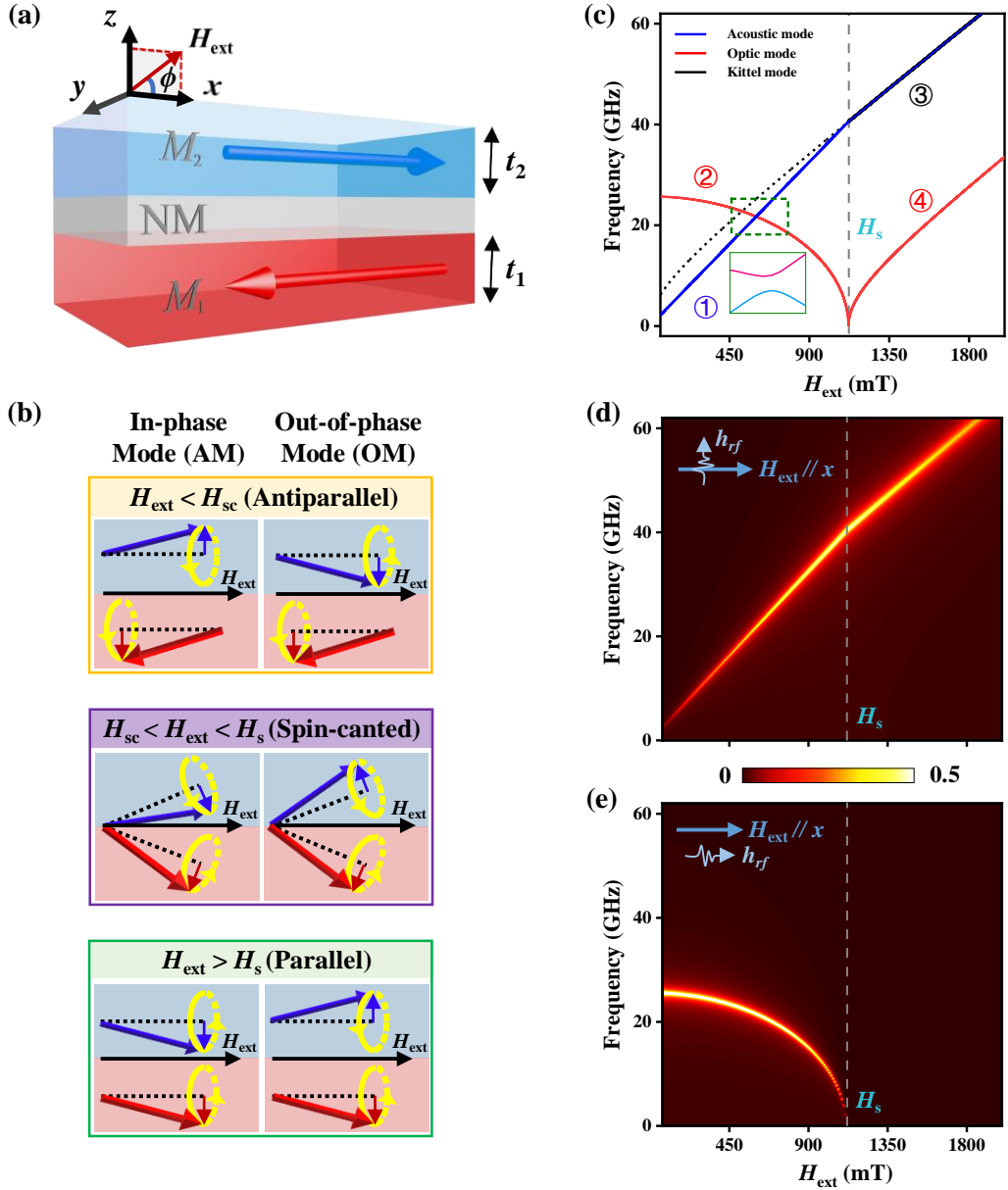

**FIG. 1.** (a) Schematic of synthetic antiferromagnet. (b) Precession-phase relations of the in-phase acoustic and out-of-phase optic modes for the magnetizations $M_1$ and $M_2$ of the two ferromagnetic layers in antiparallel (upper), spin-canted (middle), and parallel states (lower), respectively. (c) Solid lines represent theoretically calculated acoustic, optic, and Kittel modes as a function of the bias field $H_{ext}$ along $x$-axis, according to Eqs. (1), (2), and (5). Inset shows the expected "anti-crossing" at the crossing point of acoustic and optic modes as indicated by a dashed rectangle. Color plots of $f$-$H_{ext}$ dispersion in a symmetrical SAF with $t_1 = t_2 = 0.3$ nm and $J_{IEC} = -0.3 \times 10^{-4}$ J/m$^2$ under the transverse (d) and the longitudinal (e) configurations. Vertical dashed lines indicate the saturation field $H_s$.

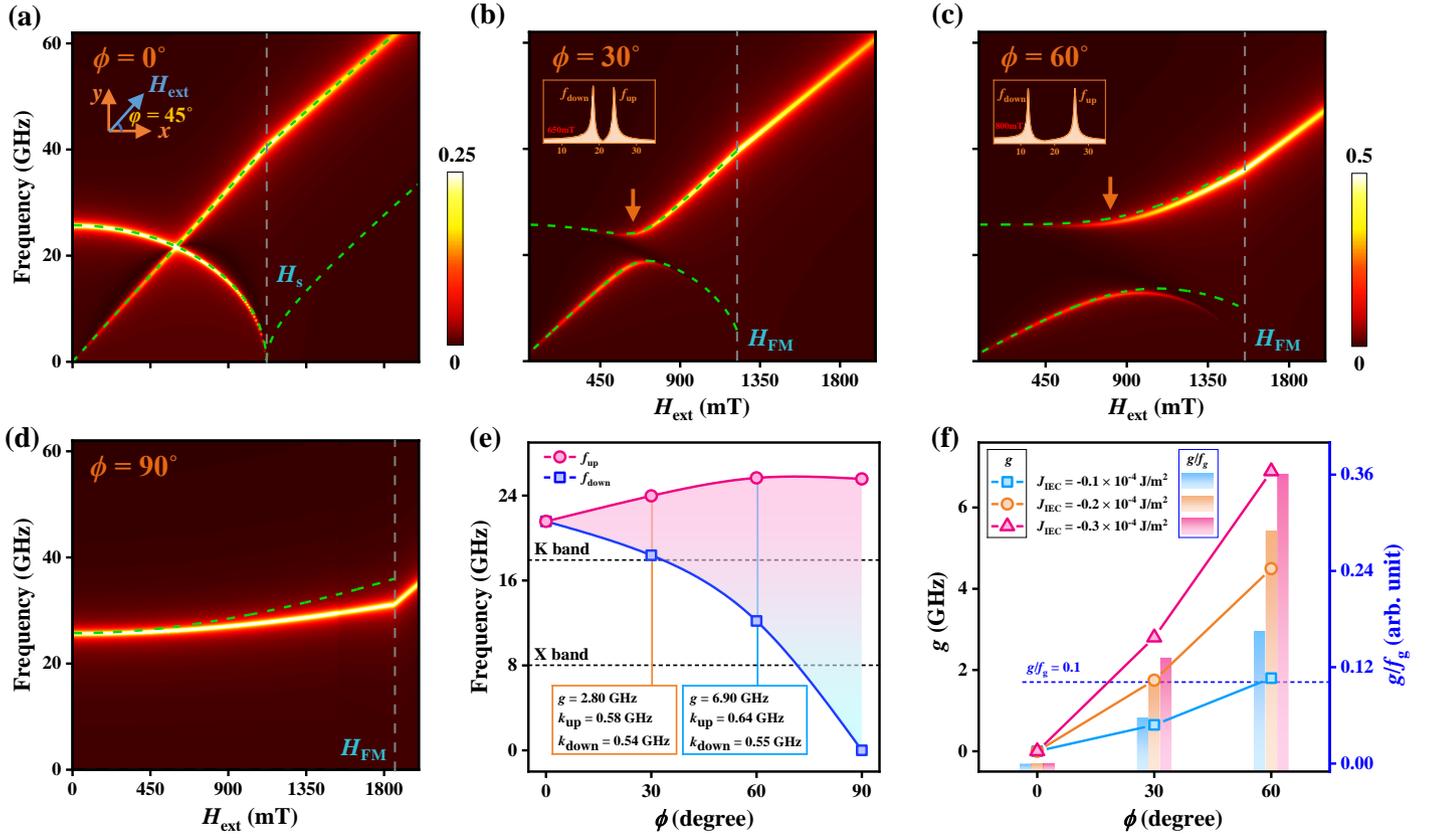

**FIG. 2.** (a)-(d) Color plots of $f$-$H_{ext}$ dispersion in a symmetrical SAF with $t_1 = t_2 = 0.3$ nm and $J_{IEC} = -0.3 \times 10^{-4}$ J/m$^2$ for $h_{rf}$ along $y$-axis and $H_{ext}$ applied at $\phi = 0°(\varphi = 45°)$, 30°, 60°, and 90°, respectively. Green dashed lines in (a) and (b)-(d) represent the theoretical calculations according to the Eqs. (1), (2) and Eq. (6), respectively. Insets in (b) and (c) are the magnon spectra at the coupling fields $H_g$ as labelled by arrows. Vertical dashed lines represent the saturation fields $H_s$ or $H_{FM}$. (e) Frequencies of the up and down modes at $H_g$ as a function of $\phi$. (f) Coupling strength $g$ (left axis, symbol + solid lines) and coupling rate $g/f_g$ (right axis, column charts) as a function of $\phi$ for three different $J_{IEC}$.

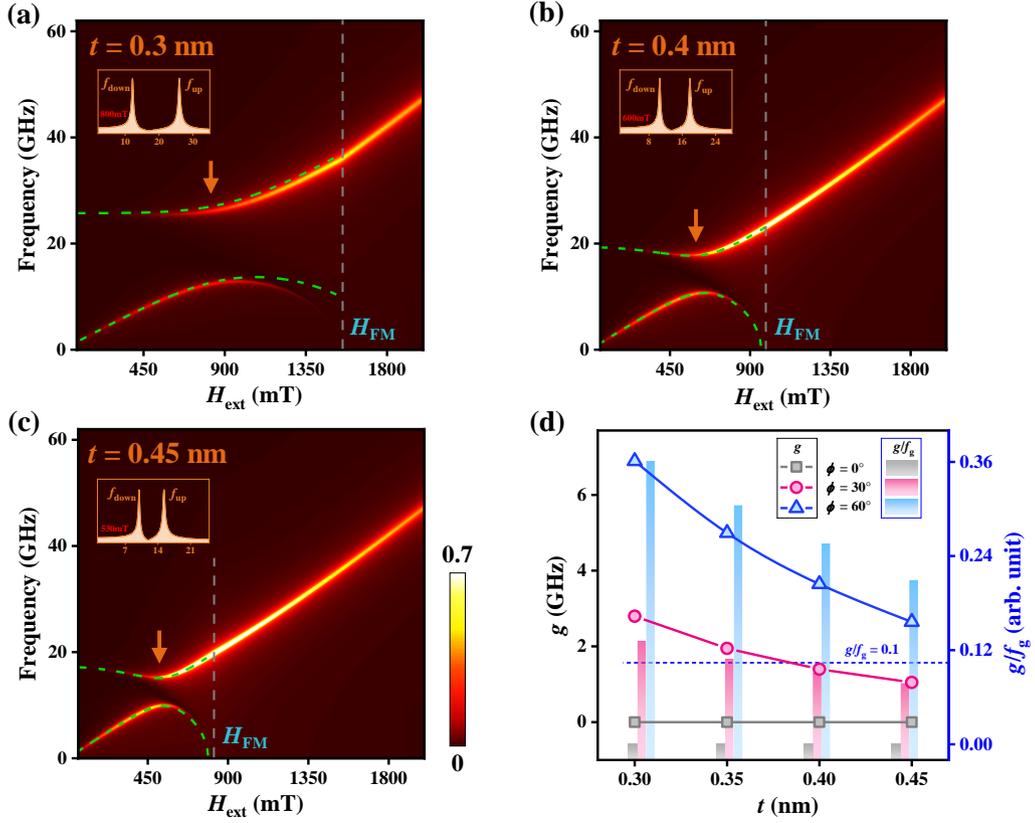

**FIG. 3.** (a)-(c) Color plots of $f$-$H_{ext}$ dispersion in symmetrical SAFs with $J_{IEC} = -0.3 \times 10^{-4}$ J/m$^2$, $\phi = 60°$ and $h_{rf}$ applied along the $y$-axis for $t_1 = t_2 = t = 0.3$, 0.4, and 0.45 nm, respectively. Green dashed lines represent theoretical calculations according to Eq. (6). Insets reveal the magnon spectra at the coupling fields $H_g$ as labelled by arrows. Vertical dashed lines reveal the saturation fields $H_{FM}$. (d) Coupling strength $g$ (left axis, symbol + solid lines) and coupling rate $g/f_g$ (right axis, column charts) as a function of thickness $t$ ($t_1 = t_2 = t$) for three different $\phi$.

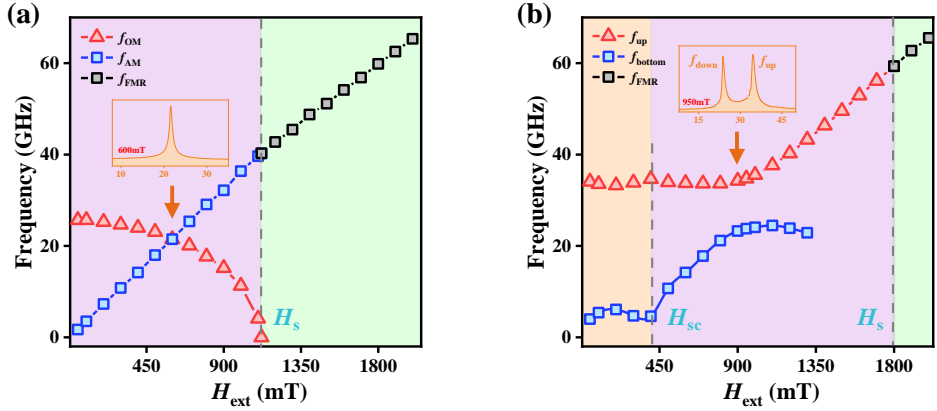

**FIG. 4.** The *f*-$H_{ext}$ dispersions for (a) symmetrical SAF with $t_1 = t_2 = 0.3$ nm, $J_{IEC} = -0.3 \times 10^{-4}$ J/m$^2$ and $\phi =$ 0° ($\varphi = 45°$) and (b) asymmetrical SAF with $t_1 = 0.45$ nm, $t_2 = 0.3$ nm, $J_{IEC} = -0.3 \times 10^{-4}$ J/m$^2$ and $\phi = 0°$ ($\varphi = 0°$) under transverse configuration. Insets represent the magnon spectra at the coupling field $H_g$ as labelled by arrows. Vertical dashed lines indicated the critical magnetic fields. The orange, purple, and green regions represent three magnetization configurations: antiparallel, spin-canted, and parallel states.